\documentclass{ws-procs9x6}

\usepackage{wrapfig}

\begin{document}

\title{$S=0$ pseudoscalar meson photoproduction from the proton.}

\author{M. Dugger\footnote{\uppercase{W}ork 
supported by the \uppercase{N}ational \uppercase{S}cience \uppercase{F}oundation.} \ and the CLAS collaboration}

\address{Arizona State University, \\
Tempe, AZ 85287-1504, USA \\
E-mail: dugger@jlab.org}

\maketitle

\abstracts{
Many measurements of pseudoscalar mesons with $S = 0$  
photoproduced on the proton have been made recently. These new data are 
particularly useful in theoretical investigations of nucleon resonances. 
How the new data from various labs complement each other 
and help fill in the gaps in the world data set is disscussed, with 
a glance at measurements to be made in the near future. Some theoretical 
techniques used to explain the data are briefly described.
}

\section{Motivation}
The $S=0$ pseudoscalar mesons include the
$\pi^0$, $\pi^+$, $\eta$, and $\eta{'}$. The pions, as the
lightest mesons, are copiously produced in the strong interaction.
Pion photoproduction data have been vital to gain a first 
glimpse of the nucleon resonance spectrum.
Whereas the pions have isospin 1,
the $\eta$ and $\eta{'}$ mesons have
isospin 0, so
resonances decaying by emitting $\eta$ or $\eta{'}$ mesons can only have isospin
$\frac{1}{2}$. Thus, $\eta$ and $\eta{'}$ mesons act as ``isospin 
filters'' for the nucleon resonance spectrum. 
Further, as the only isosinglet, $\eta{'}$ can be used 
to indirectly probe gluonic coupling to the 
proton through the flavor-singlet Goldberger-Treiman relation\cite{bass}:
\begin{equation}
g^0_A = \frac{3}{4} \frac{F_0}{m_N} (g_{\eta{'}NN} - g_{GNN}), \label{eq:ga0}
\end{equation}
where $m_N$ is the mass of the nucleon, $g_{\eta{'}NN}$ is the 
$\eta{'}$-nucleon-nucleon coupling, $g_{GNN}$ is the 
gluon-nucleon-nucleon coupling, and $F_0$ is a renormalization
constant. The flavor singlet axial charge of the nucleon ($g^0_A$) has been 
measured\cite{emc} with a value of $g^0_A = 0.20 \pm 0.35$;
however, the quark and gluon components have yet to be specifically determined.



In addition to these theoretical motivations for studying 
photoproduction of $S=0$ pseudoscalar mesons from the proton,
there are practical motivations. Electromagnetic interactions are
well understood and real photons are particularly simple (only two 
polarization states), so data from real photon beams are easier to 
analyze than data from other probes. Moreover, the $S=0$ 
two-body final states for
$\gamma p \rightarrow p X$ ($X$ is a meson), and 
$\gamma p \rightarrow n \pi^+$ have the 
benefit of having an outgoing proton or pion easy to identify and with 
relatively little contamination; $S\neq 0$ final states require
identification of a kaon usually contaminated with
both pions and protons. Thus, photoproduction of $S=0$ mesons
offers a relatively simple experimental means to explore
nucleon resonances.

\section{Resent results:}
Four experimental facilities are providing new data
on meson photoproduction from the proton: GRAAL\cite{graalNIM}, SAPHIR\cite{saphirNIM},
CB-ELSA\cite{cb-elsaNIM}, and CLAS\cite{clasNIM}. Their contributions to $S=0$ 
pseudoscalar meson photoproductions data will be summarized here.

\begin{wrapfigure}{r}{6.1cm}
\includegraphics[trim=0mm 0mm 0mm 0mm,width=0.54\textwidth]{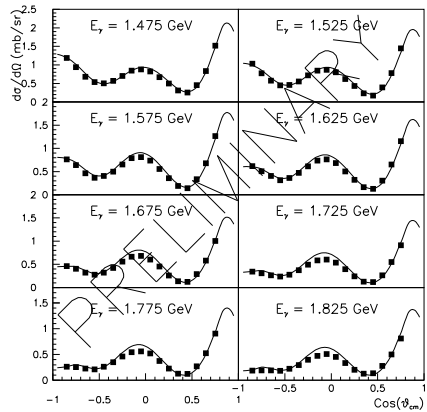}
\caption{A sample of differential cross sections for $\gamma p \rightarrow p \pi^0$.
The data points are from CLAS and the lines are from SAID.  \label{pi0_exp}} 
\end{wrapfigure}

\noindent{\textbf{a) \underline{\boldmath$\gamma p \rightarrow p \pi^0$}:}}
Prior to 2005, the world data set (compiled in SAID\cite{said})
for $\gamma p \rightarrow p \pi^0$ differential cross sections
had good coverage in incident energy ($E_\gamma$) and angle only up to 
$E_\gamma \approx 1.5$ GeV. In 2005, CB-ELSA published
results\cite{cb_pi0} which
extended coverage in $E_\gamma$ up to 3.0 GeV. While the CB-ELSA data greatly
enhance the coverage of $d\sigma/d\Omega$ for $\pi^0$ photoproduction from the proton, the 
systematic errors of the absolute normalization are estimated to be $\sim 15\%$ for $E_\gamma$ 
above 1.3 GeV. New preliminary data from CLAS  
cover $E_\gamma$ up to 2.125 GeV with an estimated systematic uncertainty
in the absolute normalization to be $< 5\%$; a sample of this data is shown in Fig.\ \ref{pi0_exp}.
An additional data set has come from GRAAL\cite{graal_private}, 
which measured differential cross sections and beam asymmetry $\Sigma$ 
for $E_\gamma$ up to 1.496 GeV.

Prior to the GRAAL measurements, 
angular coverage for $\Sigma$ was heavily biased in the
forward direction. The new GRAAL data for $\Sigma$ populate the angular range 
much more uniformly for $E_\gamma$ up to 1.496 GeV.
The rest of the polarization observables in the database are 
rather sparse. In the near future, a new generation of experiments 
specifically dedicated to polarization measurements should significantly 
expand our knowledge of polarization observables.

Thus, the world database for $\gamma p \rightarrow p \pi^0$ differential cross sections 
is becoming quite thorough
for $E_\gamma$ up to $\sim$ 3 GeV, and with coverage 
by more than one data set
up to 2.125 GeV. The new GRAAL results for beam polarization
extend the database for $\Sigma$ to $E_\gamma$ up to
1.496 GeV. All other polarization observables are very sparsely covered
in energy and angle. Future experiments are expected to enhance our knowledge 
of the polarization observables. In particular, an approved experiment\cite{double_pol_pi0}
at Jefferson Lab could start taking data for double polarization observables 
(beam and target) as soon as Fall 2006.

\begin{wrapfigure}{r}{6.1cm}
\includegraphics[trim=0mm 0mm 0mm 0mm,width=0.54\textwidth]{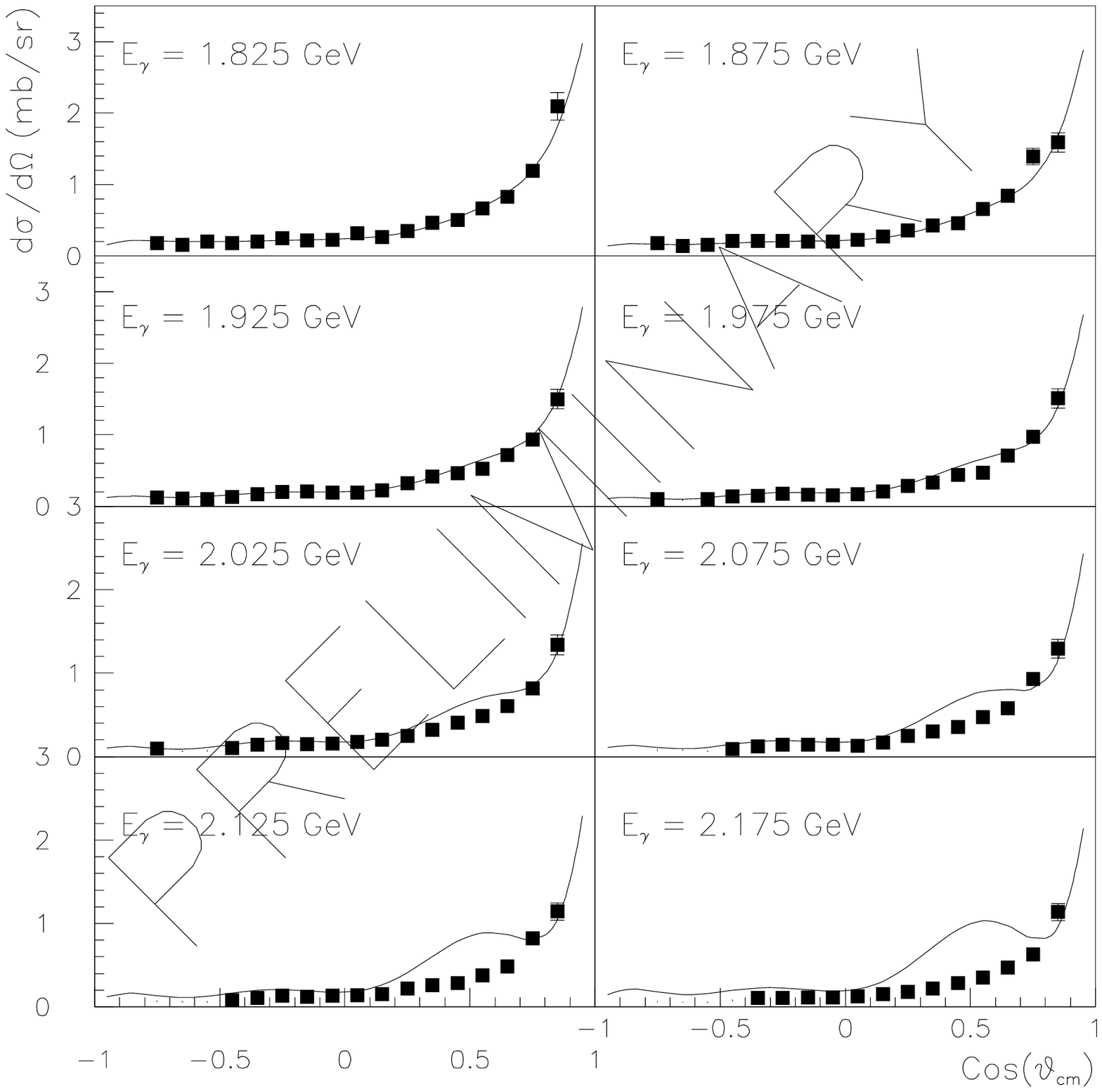}
\caption{A sample of Differential cross sections for $\gamma p \rightarrow n \pi^+$.
The data points are from CLAS and the lines are from SAID.  \label{exp}} 
\end{wrapfigure}

\noindent{\textbf {b) \underline{\boldmath$\gamma p \rightarrow n \pi^+$}:}}
Of the four collaborations mentioned above, only the CLAS Collaboration
has an analysis effort underway for the reaction $\gamma p \rightarrow n \pi^+$
which is in a very preliminary state.
Preliminary differential cross section results for $\gamma p \rightarrow n \pi^+$
agree well with the SAID parameterization for $E_\gamma$ up to 1.925 GeV
(see Fig.\ \ref{exp}).

The CLAS cross sections were measured for $E_\gamma$ up to 2.225 GeV, and for
$E_\gamma$ above 1.925 GeV the world database is nearly non-existent for the
central angles between 50 and 150 degrees. It is here that the CLAS results 
will be of use in determining the differential cross sections for this reaction.

The coverages in the world database for polarization observables for the 
$\gamma p \rightarrow n \pi^+$ reaction are in a comparable state as that
for the $\pi^0$ reaction, with the exception that the beam polarization observables
are not as weighted in the forward direction. As with the $\pi^0$ reaction,
there is an approved experiment\cite{double_pol_pi0} at Jefferson Lab to obtain  
double polarization observables that could start taking data as soon as
Fall 2006.

\noindent{\textbf{c) \underline{\boldmath$\gamma p \rightarrow p \eta$}:}}
Before 2002 the world database for $\gamma p \rightarrow p \eta$ differential cross sections 
was only well covered for $E_\gamma$ from threshold 
(0.707 GeV) up to 0.8 GeV. In 2002 GRAAL published results\cite{graal_eta} 
on $d\sigma/d\Omega$ for $E_\gamma$ up to 1.1 GeV, and  CLAS 
published\cite{clas_eta} $d\sigma/d\Omega$ for $E_\gamma$ up to 1.95 GeV.
More recently (2005), CB-ELSA published\cite{cb_elsa_eta} $d\sigma/d\Omega$ results for 
$E_\gamma$ up to 3 GeV.

In 1998, GRAAL published\cite{graal_pol}  $\gamma p \rightarrow p \eta$ beam polarization results 
for $E_\gamma$ up to 1.45 GeV. Polarization observables for $\gamma p \rightarrow p \pi^+$ remain
very sparsely populated. 
This past summer, data were taken at CLAS for beam polarization that should allow extraction
of that observable for $\eta$ photoproduction for $E_\gamma$ up to 2.1 GeV. As with the pions, 
an approved 
experiment\cite{double_pol_eta} at Jefferson Lab should 
help fill in the database for single and double polarization observables 
in $\eta$ photoproduction from the proton.

\begin{wrapfigure}{r}{6.1cm}
\includegraphics[trim=0mm 0mm 0mm 0mm,width=0.54\textwidth]{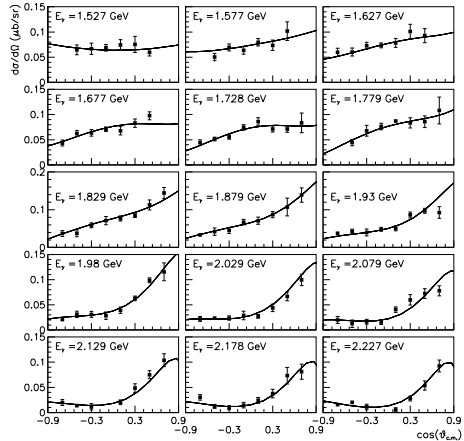}
\caption{Differential cross sections for $\gamma p \rightarrow p \eta{'}$.
The data points are from CLAS and the lines are from the HK model described in the text.
\label{etap}} 
\end{wrapfigure}

\noindent{\textbf{d) \underline{\boldmath$\gamma p \rightarrow p \eta{'}$}:}}
Prior to 1998, only 18 $\eta{'}$ photoproduction events had been
measured (11 events from the ABBHHM bubble chamber experiment\cite{abbhhm}, 
and 7 events from the AHHM streamer
chamber experiment\cite{ahhm}).  In 1998, the SAPHIR
collaboration published results\cite{saphir} extracted from an
additional 250 $\eta{'}$ exclusive events.
By contrast, the new (unpublished) CLAS results have over 2$\times
10^5$ $\eta{'}$ photoproduction events detected and used to extract
differential cross sections shown in  Fig.\ \ref{etap}. These CLAS results span $E_\gamma$ 
from 1.527 to 2.227 GeV.
No polarization observables have been measured for this reaction.
Therefore, the differential cross sections provide the only 
experimental data for the reaction $\gamma p \rightarrow p \eta{'}$.

\section{Theoretical Results}
As noted above, there are many new differential cross section data
for the reactions discussed here. However, these 
alone are not sufficient to constrain theoretical 
models to the extent that resonances can be uniquely determined. 
More data on the polarization observables are desperately needed, and
a coupled channel approach is required, 
in order to constrain the contributions of various resonances.

One step in this direction comes from a model\cite{cb_pi0,cb_elsa_eta,ak} developed by A. V. Anisovich, 
E. Klempt, A.Sarantsev, and U.Thoma (AKST model) that couples the 
reactions $\gamma p \rightarrow p \pi^0$, $n \pi^+$, and $p \eta$. AKST included
published differential cross sections, as well as the recent GRAAL beam polarization observables. The
model uses a $K$ matrix approach for the $S_{11}(1535)$ and the $S_{11}(1650)$ resonances. 
The remaining resonances are described by Breit-Wigner amplitudes. The model also includes 
reggeized $u-$ and $t-$channel contributions. Results from their analysis\cite{cb_pi0} find
evidence for a previously unseen $D_{15}(2070)$ resonance, and indications 
for a new $P_{13}(2200)$ resonance.

One model that
considers the $\eta{'}$ exclusively comes from K. Nakayama and H. Haberzettl (NH). 
This model is based upon a relativistic
meson-exchange model of hadronic interactions.
Allowed processes include $s$-, $t$-, and $u$-channel contributions.
The intermediate mesons in the $t-$channel exchanges
are the $\omega$ and $\rho^0$. The NH model includes the
$S_{11}(1535)$ and $P_{11}(1710)$ resonances ($j = 1/2$), which are
known to decay strongly to the ${\eta}N$ channel\cite{PDG}, and also 
includes two additional $S_{11}$ and two additional
$P_{11}$ resonances, albeit with relatively small couplings.
The present adaptation of the NH model to the CLAS
data now also requires $j = 3/2$ resonances [$P_{13}(1940)$, $D_{13}(1780)$, and $D_{13}(2090)$].
The observed $u$-channel contribution seen here allows the $g_{\eta{'}NN}$
coupling to be extracted (albeit in a model-dependent way). The
value of $g_{\eta{'}NN}$ found from the particular NH fit is 1.33.
Since differential cross sections alone do not provide sufficient
constraints to this model, the $g_{\eta{'}NN}$ values
should be taken with caution.
Nonetheless, this value is consistent with the analysis of T. Feldmann\cite{feld} which
gives $g_{\eta{'}NN} = 1.4 \pm 1.1.$

\section{Summary}
While there has been much progress in obtaining differential cross section
data for pseudoscalar $S=0$ meson photoproduction from the proton, and some new
beam polarization ($\Sigma$) measurements for the pions and $\eta$, more 
polarization observables are needed in order to provide constraints to 
theoretical models. Experiments to obtain these needed constraints are planned 
for the near future. 
A comprehensive program for single, 
double and even triple polarization measurements in 
photoproduction is in preparation at Jefferson Lab. 
From the data already taken, there appears to be evidence for a new $D_{15}$ 
resonance at 2.09 GeV and indications of a new $P_{13}$ at 2.20 GeV. 
In addition
to improving our knowledge about resonaces and their parameters by fitting the
$\gamma p \rightarrow p \pi^0$, $n \pi^+$, and $p \eta$ data, the new CLAS 
$\eta{'}$ data also has been analyzed 
to suggest a value of 
$g_{\eta{'}NN} \approx 1.3$, consistent with theoretical predictions. 
When this value can be determined
with a high degree of confidence, it can be used to indirectly determine the gluonic
coupling to the proton through the flavor-singlet Goldberger-Treiman relation.

\end{document}